\documentclass[namedreferences]{SolarPhysics}
\usepackage[optionalrh]{spr-sola-addons} % For Solar Physics 
\usepackage{graphicx}        % For eps figures, newer & more powerfull
\usepackage{color}           % For color text: \color command
\usepackage{url}             % For breaking URLs easily trough lines
\usepackage{lscape}
\usepackage{ccaption}
\usepackage{ulem}

\newcommand{\etal}{{\it et al.}}

\newcommand{\solphys}{{\it Solar Phys.}}

\begin{document}

\begin{article}

\begin{opening}

\title{Coronal-Temperature-Diagnostic Capability of the \textit{Hinode}/\textit{X-Ray Telescope}
Based on Self-Consistent Calibration. II. Calibration with on-Orbit Data}

\author{N.~\surname{Narukage}$^{1}$\sep
        T.~\surname{Sakao}$^{1}$\sep
        R.~\surname{Kano}$^{2}$\sep
        M.~\surname{Shimojo}$^{2}$\sep
        A.~\surname{Winebarger}$^{3}$\sep
        M.~\surname{Weber}$^{4}$\sep
        K.~K.~\surname{Reeves}$^{4}$\sep
       }

\runningauthor{N. Narukage \etal}
\runningtitle{Calibration of the \textit{Hinode}/\textit{X-Ray Telescope} with on-Orbit Data}
\institute{$^{1}$ Institute of Space and Astronautical Science/\\
                  Japan Aerospace Exploration Agency (ISAS/JAXA),\\
                  3-1-1 Yoshinodai, Sagamihara, Kanagawa, 229-8510, Japan\\
                  email: \url{narukage@solar.isas.jaxa.jp}\\
           $^{2}$ National Astronomical Observatory of Japan (NAOJ),\\
                  2-21-1 Osawa, Mitaka, Tokyo, 181-8588, Japan\\
           $^{3}$ NASA/Marshall Space Flight Center,\\
                  AL 35812, USA\\
           $^{4}$ Smithsonian Astrophysical Observatory,\\
                  60 Garden Street, Cambridge, MA 02138, USA\\
           }

% ************************************************************************ ABSTRACT
\begin{abstract}
The \textit{X-Ray Telescope} (XRT) onboard the \textit{Hinode} satellite is an X-ray imager
that observes the solar corona with the capability of diagnosing coronal temperatures
from less than 1~MK to more than 10~MK.
To make full use of this capability, \citeauthor{nar11} (\solphys{} \textbf{269}, 169, \citeyear{nar11})
determined the thickness of each of the X-ray focal-plane analysis filters based on
calibration measurements from the ground-based end-to-end test.
However, in their paper, the calibration of the thicker filters for observations of active regions and flares,
namely the med-Be, med-Al, thick-Al and thick-Be filters, was insufficient due to the insufficient X-ray flux
used in the measurements.
In this work, we recalibrate those thicker filters using quiescent active region data
taken with multiple filters of XRT.
On the basis of our updated calibration results, we present the revised coronal-temperature-diagnostic
capability of XRT.
\end{abstract}
\end{opening}

% ************************************************************************ SECTION
\section{Introduction}
\label{introduction} 

The \textit{X-Ray Telescope} (XRT) \cite{gol07, kan08} onboard \textit{Hinode} \cite{kos07} is
able to perform detailed imaging observations of
a wide variety of coronal plasmas in a temperature range from below 1~MK to
well above 10~MK with a 1$''$ CCD pixel size.
One of the most significant scientific features of the XRT is its
coronal-temperature-diagnostic capability, namely its capability to make temperature maps
with up to a full-Sun field of view and a high temporal resolution ($\sim$ a few ten seconds).

In order to have XRT perform coronal-temperature diagnostics with its full capability,
\inlinecite{nar11} (hereafter, Paper~I) measured the thicknesses of the X-ray focal-plane analysis filters and
characterized the effect of on-orbit contamination.
In Paper~I, the thicknesses of the filters were calibrated using X-ray emission lines
in the ground-based end-to-end test at the X-ray Calibration Facility (XRCF) at NASA/Marshall Space Flight Center.
However, the calibration of the thicker filters
(med-Be, med-Al, thick-Al, and thick-Be filters; hereafter, thick filter group) remained less certain,
since low-energy X-rays do not have enough transmission through them.
Less than three X-ray lines were able to be used for the calibration of the thicker filter group,
while five emission lines were able to be used for the calibration of the thinner filters.
The details are described in Appendix A.2 of Paper~I.

In this paper, we refine the thickness measurements of the XRT filters with
on-orbit datasets of a quiescent active region taken with XRT.
As a result of this work,
the filter thicknesses of the thick filter group are updated
and the thicknesses of other filters ({\it i.e.}, thinner filters) are confirmed to be consistent with those in Paper~I.
On the basis of our calibration results, we update the coronal-temperature-diagnostic
capability of XRT with the filter-ratio method described in Paper~I.

In Section~\ref{sec:datasets}, the datasets used for the on-orbit calibration are described.
In Section~\ref{sec:calibration}, we explain the calibration method and show its result.
In Section~\ref{sec:confirmation}, we confirm the validity of our calibration method and result.
Finally, in Section~\ref{sec:summary},
we evaluate the coronal-temperature-diagnostic capability of XRT based on our revised calibration result.

% ************************************************************************ SECTION
\section{Datasets for Calibration of XRT Filters}
\label{sec:datasets}

In order to obtain meaningful calibration results, the following two properties are required for the X-ray source:
\begin{itemize}
\item[(\textit{i})]  The X-ray emission should be intense enough so that the signal through each filter is significant.
\item[(\textit{ii})] The X-ray emission should be stable while the data are taken.
\end{itemize}

% ........................................................................ FIGURE
\begin{figure}
\centerline{\includegraphics[width=12.0cm,clip=]{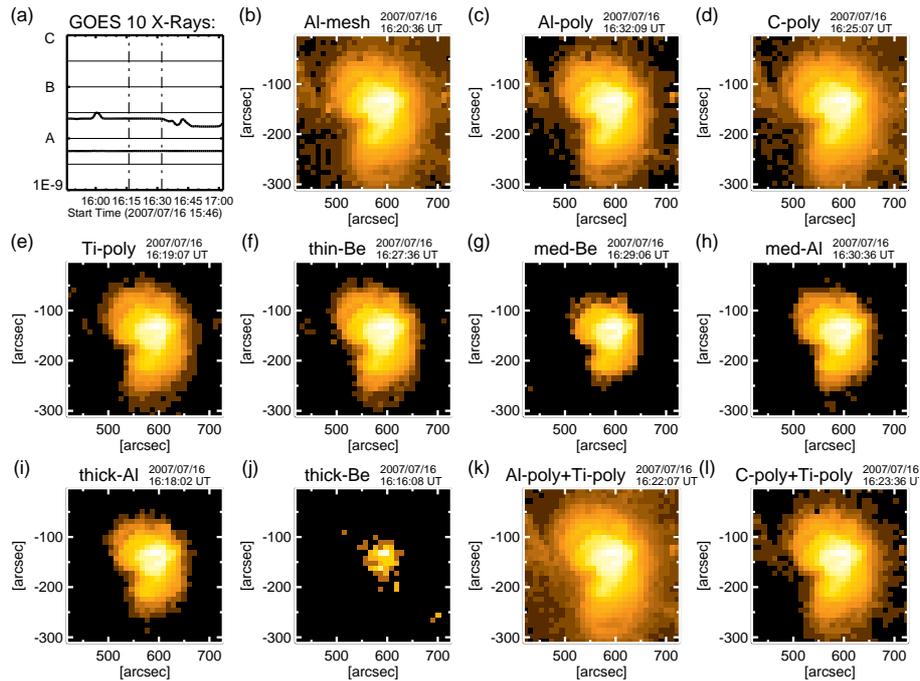}}
\vspace{2.0mm}
\caption{
Datasets of the quiescent active region (AR 10963) used for the calibration.
These images were taken by XRT with multiple filters from 16:16 to 16:32 UT on 2007 July 16.
{\bf (a)} shows the GOES X-ray intensity flux
from the first image $-30$ minutes to the last image $+30$ minutes.
XRT data were taken in the time interval between the two vertical dash-dotted lines.
{\bf (b)--(l)} display XRT images taken with different filters (or filter combinations)
that are shown at the top of each panel.
Though the original data were taken with a spatial resolution of 1$''$ (CCD pixel size),
these data are binned in $10 \times 10$~pixels to attain a good signal-to-noise ratio.
}
\label{fig:datasets}
\end{figure}

In this work, we select the quiescent active region (AR) 10963 (NOAA number), shown in Figure~\ref{fig:datasets},
as the X-ray source for the calibration.
XRT observed this active region with multiple filters around 16:15 UT on 2007 July 16.
We bin the data in 10 $\times$ 10 pixels (corresponding to 10.3$''$ $\times$ 10.3$''$)
in order to improve the signal-to-noise ratios (see Figure~\ref{fig:datasets}).
The binned data taken with the thick filter group
had a good signal quality to meet requirement~(\textit{i}).
During the observations, the GOES X-ray intensity level was stable, as shown in Figure~\ref{fig:datasets} (a),
satisfying requirement~(\textit{ii}).

% ************************************************************************ SECTION
\section{Calibration of XRT Filters}
\label{sec:calibration}

To refine the filter thickness values using only XRT data, we follow these steps:
\begin{itemize}
\item[(I)] Adopting the filter thicknesses in Paper~I for Ti-poly and thin-Be filters,
derive maps of the filter-ratio temperature, $T_\mathrm{ratio}^\mathrm{Be/Ti}$, and volume emission measure,
${VEM}_\mathrm{ratio}^\mathrm{Be/Ti}$, for the quiescent active region (AR 10963).
\item[(II)] Estimate the expected X-ray intensity, $I_\mathrm{exp}$,
from AR 10963 as a function of filter thickness
based on the values derived in step~(I).
\item[(III)] Calibrate the thicknesses of the other filters to match $I_\mathrm{exp}$
with the observed intensity, $I_\mathrm{obs}$, for AR 10963.
\end{itemize}

We note that an active region should contain multi-temperature structures,
and, in the above steps, we assume a single temperature and volume emission
measure derived with the filter-ratio method. In Section~5.3 of Paper~I,
we found that the filter-ratio temperature derived with XRT is the
DEM-weighted (DEM, differential emission measure) mean temperature. As shown in Figure~17 of Paper~I, for an
active region, we saw that 16 different filter pairs, including all XRT filters
except thick-Al and thick-Be, give temperatures comparable to the
DEM-weighted mean temperature. Hence, the DEM-weighted mean temperature can be used to derive
the expected X-ray intensity to be observed with the XRT
filters, except for thick-Al and thick-Be.
Whether $T_\mathrm{ratio}^\mathrm{Be/Ti}$ can be also used to derive the expected X-ray intensity
with thick-Al and thick-Be filters remains unknown. Hence, the validity of
this method to derive the thick-Al and thick-Be filter thicknesses is
discussed in Section~\ref{sec:confirmation}.

The filter-ratio method is described in Section~5.1 of Paper~I,
where the X-ray intensity, $I$, is written as $DN / t$, where DN stands for data numbers.

% ........................................................................ FIGURE
\begin{figure}
\centerline{\includegraphics[width=12.0cm,clip=]{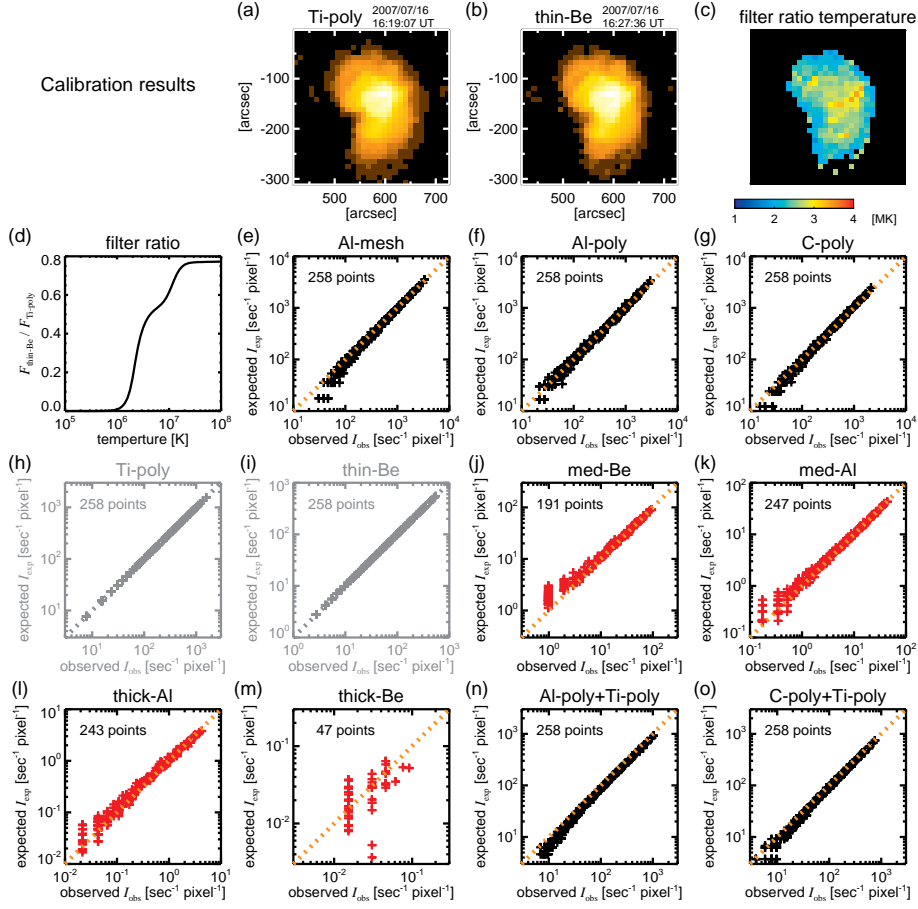}}
\caption{
Calibration results using $T_\mathrm{ratio}^\mathrm{Be/Ti}$ and ${VEM}_\mathrm{ratio}^\mathrm{Be/Ti}$ derived from
Ti-poly and thin-Be filter-pair data of AR 10963.
{\bf (a)} and {\bf (b)} are the observed X-ray images and {\bf (c)} is the derived filter-ratio
$T_\mathrm{ratio}^\mathrm{Be/Ti}$ from the data sets in {\bf (a)} and {\bf (b)}, and the function of filter-ratio shown in {\bf (d)}.
The black and red plus signs in {\bf (e)--(o}) are the data points
based on the filter thicknesses determined in Paper~I and in this paper, respectively.
Each plus sign is determined for a binned area ($10 \times 10$~pixels area)
where $T_\mathrm{ratio}^\mathrm{Be/Ti}$ and the X-ray emission can be obtained with certainty.
The number of plus signs is indicated in each panel.
The lines depicted by orange (gray) bars in {\bf (e)--(g)} and {\bf (j)--(o)} ({\bf (h)--(i)}) indicate the relation of
$I_\mathrm{exp} = I_\mathrm{obs}$.
}
\label{fig:cal result}
\end{figure}

% ........................................................................ FIGURE
\begin{figure}
\centerline{\includegraphics[width=12.0cm,clip=]{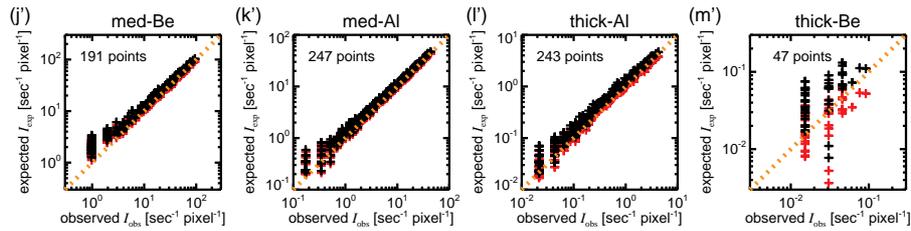}}
\caption{
{\bf (j')--(m')} Comparison between $I_\mathrm{obs}$ and $I_\mathrm{exp}$ based on the filter thicknesses
determined in Paper~I, for the thick filter group, shown with black plus signs.
As a reference, the data points based on the thicknesses updated in this paper are shown with red plus signs,
which are overlaid with the black signs.
The number of plus signs is indicated in each panel and the lines depicted by orange bars
represent the same as in Figures~\ref{fig:cal result} (e)--(g) and (j)--(o).
}
\label{fig:Paper I result}
\end{figure}

We take $T_\mathrm{ratio}^\mathrm{Be/Ti}$ (see Figure~\ref{fig:cal result} (c)) and
${VEM}_\mathrm{ratio}^\mathrm{Be/Ti}$ estimated using Ti-poly and thin-Be filters
as a ``standard candle" in this analysis due to the following reasons.
First, these filters were well calibrated in Paper~I.
Hence, the temperature response profiles, $F(T)$ (see Equation~(39) in Appendix~D.2 of Paper~I for details),
for these filters can be regarded as highly reliable.
Second, the ratio of $F(T)$ between Ti-poly and thin-Be filters increases monotonically
in a temperature range from about 1~MK to more than 10~MK (see Figure~\ref{fig:cal result} (d)), {\it i.e.},
this filter pair can determine the filter-ratio temperature uniquely in a broad temperature range.

The black plus signs in Figures~\ref{fig:cal result} (e)--(g), (n), and (o) and
in Figures~\ref{fig:Paper I result} (j')--(m') show the relation between $I_\mathrm{obs}$ and $I_\mathrm{exp}$ using
the filter thicknesses determined in Paper~I.
In the case of the thinner filters, namely, Al-mesh, Al-poly and C-poly filters (see Figures~\ref{fig:cal result} (e)--(g)),
the black plus signs follow nicely the line, depicted by orange bars, where $I_\mathrm{exp} = I_\mathrm{obs}$.
This implies that the thicknesses of the thinner filters are indeed well calibrated in Paper~I.
Moreover, stacks of thinner filters and Ti-poly filter ({\it e.g.}, Al-poly+Ti-poly and C-poly+Ti-poly)
also give an $I_\mathrm{exp}$ consistent with $I_\mathrm{obs}$, as shown in Figures~\ref{fig:cal result} (n) and (o).
Considering the above consistency, it is also natural to expect that
the thicknesses of Ti-poly and thin-Be filters calibrated in Paper~I and used to derive
$T_\mathrm{ratio}^\mathrm{Be/Ti}$ and ${VEM}_\mathrm{ratio}^\mathrm{Be/Ti}$ should be reliable, as well.
It is also evident in Figures~\ref{fig:cal result} (e)--(g), and (j)--(o) that the plus sings appear more scattered
for the lower $I_\mathrm{obs}$ values; this is caused by the photon noise.
Furthermore, since this calibration is performed with $T_\mathrm{ratio}^\mathrm{Be/Ti}$ and
${VEM}_\mathrm{ratio}^\mathrm{Be/Ti}$ derived from the data taken with Ti-poly and thin-Be filters,
$I_\mathrm{exp}$ should perfectly match $I_\mathrm{obs}$ for these filters (see Figures~\ref{fig:cal result} (h)--(i)).
On the other hand, the black plus signs for the thick filter group
(med-Al, med-Be, thick-Al, and thick-Be filters) deviate from the line depicted by orange bars
(see Figures~\ref{fig:Paper I result} (j')--(m')).
This means that the thick filter group is not well calibrated, as was pointed out in Paper~I.

% ........................................................................ TABLE
\begin{landscape}
\begin{table}
\caption{Calibrated Thicknesses of Focal-Plane Analysis Filters.}
\label{tbl:FPAF}
\begin{tabular}{ccccccc}
\hline
FW-pos$^{(a)}$ & filter name    & \multicolumn{4}{c}{calibrated values}                                           \\
\cline{3-6}
               &                & metal thickness         & on-orbit           & on-orbit               &                           \\
               &                & at fabrication          & pure metal         & oxidized layer         & support                   \\
               &                &                         & thickness          & thickness              &                           \\
\hline
 1-0           & open           & --                      & --                 & --                     & --                        \\
 1-1           & Al-poly        & Al$^{(b)}$  1470 {\AA}  & Al     1412 {\AA}  &  Al$_2$O$_3$  75 {\AA} & poly$^{(c)}$ 2656 {\AA}   \\
 1-2           & C-poly         & C$^{(d)}$   5190 {\AA}  & C      5190 {\AA}  &  --                    & poly 3478 {\AA}           \\
 1-3           & thin-Be        & Be$^{(e)}$ 10.47 $\mu$m & Be    10.46 $\mu$m &  BeO         150 {\AA} & --                        \\
 1-4           & {\bf med-Be}   & Be          28.5 $\mu$m & Be    28.49 $\mu$m &  BeO         150 {\AA} & --                        \\
 1-5           & {\bf med-Al}   & Al          12.7 $\mu$m & Al    12.69 $\mu$m &  Al$_2$O$_3$ 150 {\AA} & --                        \\
 2-0           & open           & --                      & --                 & --                     & --                        \\
 2-1           & Al-mesh        & Al          1700 {\AA}  & Al     1583 {\AA}  &  Al$_2$O$_3$ 150 {\AA} & 77\% trans. mesh$^{(f)}$  \\
 2-2           & Ti-poly        & Ti$^{(g)}$  2380 {\AA}  & Ti     2338 {\AA}  &  TiO$_2$      75 {\AA} & poly 2522 {\AA}           \\
 2-3           & G-band         & --                      & --                 & --                     & --                        \\
 2-4           & {\bf thick-Al} & Al          27.5 $\mu$m & Al    27.49 $\mu$m &  Al$_2$O$_3$ 150 {\AA} & --                        \\
 2-5           & {\bf thick-Be} & Be           295 $\mu$m & Be   294.99 $\mu$m &  BeO         150 {\AA} & --                        \\
\hline
\end{tabular}
\begin{description}
\item[NOTE --] The filter names whose filter thicknesses are updated from Paper~I are shown in bold font.\\
The metal thicknesses at fabrication determined in Paper~I are as follows:\\
med-Be: 26.90~$\mu$m, med-Al: 12.26~$\mu$m, thick-Al: 26.1~$\mu$m, thick-Be: 252.8~$\mu$m
\item[$^{(a)}$] ``FW" and ``pos" indicate the filter wheel number and position on the filter wheel, respectively.
\item[$^{(b)-(e), (g)}$] ``Al" means aluminum, ``poly" polyimide (C$_{22}$H$_{10}$N$_{2}$O$_{5}$), ``C" carbon, ``Be" beryllium, and\\
                         ``Ti" titanium.
\item[$^{(f)}$] The mesh for the Al-mesh filter is made of stainless steel.
\end{description}
\end{table}
\end{landscape}

After the previous steps, in step~(III), we derive the adjusted metal thicknesses ``at fabrication",
with a resolution of 0.1~$\mu$m, that provide the best fits of $I_\mathrm{exp}$ to $I_\mathrm{obs}$,
as shown by the red plus signs in Figures~\ref{fig:cal result} (j)--(m).
Thicknesses thus derived are adopted as the newly calibrated filter thicknesses.
The calibrated filter thicknesses are summarized in Table~\ref{tbl:FPAF}.

For coronal-temperature diagnostics, the on-orbit thickness of pure metal and the oxidized layer
for each filter should be calibrated, instead of the metal thickness at fabrication.
Thus, we incorporate the effects of the oxidized layer and the contamination layer
on the focal-plane analysis filters using the values obtained in Paper~I,
since such values were deduced from the well-calibrated thinner filters (see Appendix B.3.3 in Paper~I).
The relation among the metal thickness at fabrication, the on-orbit pure metal thickness,
and the on-orbit oxidized layer thickness, summarized in Table~\ref{tbl:FPAF}, is explained
by Equation~(56) in Appendix~I of Paper~I.
The effect of the contamination accumulated on the CCD that was calibrated in Paper~I
is also taken into account in this calibration.

% ........................................................................ FIGURE
\begin{figure}
\centerline{\includegraphics[width=10.0cm,clip=]{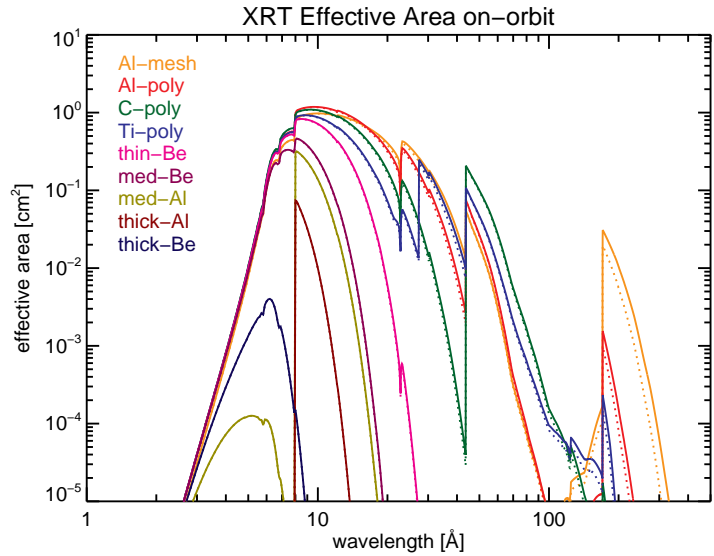}}
\caption{
Calibrated effective area of XRT.
The laminar contaminant accumulated on the focal-plane analysis filters and the CCD are considered.
The solid and dotted lines show the effective area just after CCD bakeout and one month after the bakeout, respectively.
We assume that 800~{\AA} of contaminant accumulated on CCD one month after the bakeout.
}
\label{fig:eff area}
\end{figure}

% ........................................................................ FIGURE
\begin{figure}
\centerline{\includegraphics[width=10.0cm,clip=]{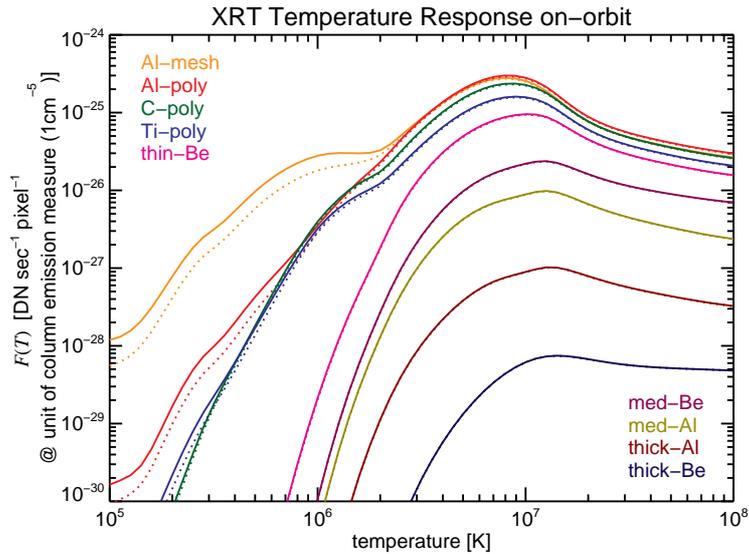}}
\caption{
Calibrated response of XRT to the coronal temperature, $F(T)$.
The laminar contaminant accumulated on the focal-plane analysis filters and the CCD are considered.
The solid and dotted lines show the temperature response just after the CCD bakeout and one month after the bakeout, respectively.
We assume that 800~{\AA} of contaminant accumulated on the CCD one month after the bakeout.
}
\label{fig:response}
\end{figure}

As a result of the analysis described above, we confirm that
the thicknesses of Al-mesh, Al-poly, C-poly, Ti-poly, and thin-Be filters were well calibrated in Paper~I,
while we updated the filter thicknesses of med-Al, med-Be, thick-Al, and thick-Be filters.
The effective area and temperature response using the parameters in Table~\ref{tbl:FPAF}
are shown in Figures~\ref{fig:eff area} and \ref{fig:response}, respectively.

Finally, we assess the potential error introduced by using a 0.1~$\mu$m resolution in this calibration.
An aluminum filter of 0.1~$\mu$m thickness has a transmissivity of about 99~\% at a wavelength of around 8~{\AA}
\cite{hen93} where the effective area of the med-Al and thick-Al filters has a peak (see Figure~\ref{fig:eff area}).
A beryllium filter of 0.1~$\mu$m thickness has a transmissivity of more than 99.4~\%
at a wavelength of less than 10~{\AA} \cite{hen93},
where the peak of the effective area for the med-Be and thick-Be filters is located (see Figure~\ref{fig:eff area}).
Hence, a resolution of 0.1~$\mu$m thickness causes a maximum error of 1~\%
in the X-ray intensity detected with the thick filter group.
Since this error is less than the photon noise of typical data sets for the coronal-temperature analysis ($\gtrsim$ 3~\%),
a resolution of 0.1~$\mu$m thickness is enough for coronal-temperature diagnostics.

% ************************************************************************ SECTION
\section{Confirmation of Calibration Results}
\label{sec:confirmation}

In this section, we review the validity of our method considering two approaches and
using the new set of the filter thicknesses, tabulated in Table~\ref{tbl:FPAF}, as follows:
\begin{itemize}
\item Using the DEM of the active region, derived from the data sets taken by
\textit{Hinode}/\textit{EUV Imaging Spectrometer} (EIS) \cite{cul07} and XRT,
we check whether the quantity $I_\mathrm{exp}$ represents the expected X-ray intensity from the active region
(Section~\ref{sec:confirmationA}).
\item Using the filter-ratio temperatures, we check whether the newly-obtained filter thicknesses can also give
self-consistent results with a quiescent active region other than the one used in Section~\ref{sec:calibration},
{\it i.e.}, with a different thermal structure (Section~\ref{sec:confirmationB}).
\end{itemize}

% ************************************************************************ SUB-SECTION
\subsection{Confirmation Using the DEM of the Active Region Derived from EIS Data}
\label{sec:confirmationA}

% ........................................................................ FIGURE
\begin{figure}
\centerline{\includegraphics[width=10.0cm,clip=]{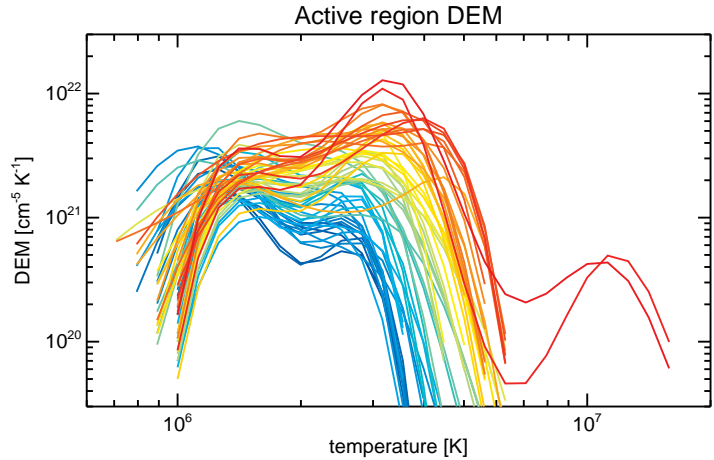}}
\caption{
The DEM of AR 10963 derived from EIS and XRT data.
The EIS data were obtained around 14:55 UT on 2007 July 16, and
the XRT data is the same as in Figure~\ref{fig:datasets}.
Each colored line shows the DEM of a region indicated by each colored box
in Figure~\ref{fig:confirmationA} (a).
}
\label{fig:DEM}
\end{figure}

% ........................................................................ TABLE
\begin{table}
\caption{Dataset used for DEM Analysis.}
\label{tbl:DEM}
\begin{tabular}{ccc}
\hline
N$^0$ & line or filter      & instrument \\
\hline
1  & Fe X 184.536 \AA    & EIS \\
2  & Fe XI 188.216 \AA   & EIS \\
3  & Fe XI 192.813 \AA   & EIS \\
4  & Fe XII 195.119 \AA  & EIS \\
5  & Fe XIII 203.826 \AA & EIS \\
6  & Fe XV 284.160 \AA   & EIS \\
7  & Fe XVI 262.984 \AA  & EIS \\
8  & Open/Ti-poly        & XRT \\
9  & Open/thin-Be        & XRT \\
\hline
\end{tabular}
\end{table}

% ........................................................................ FIGURE
\begin{figure}
\centerline{\includegraphics[width=12.0cm,clip=]{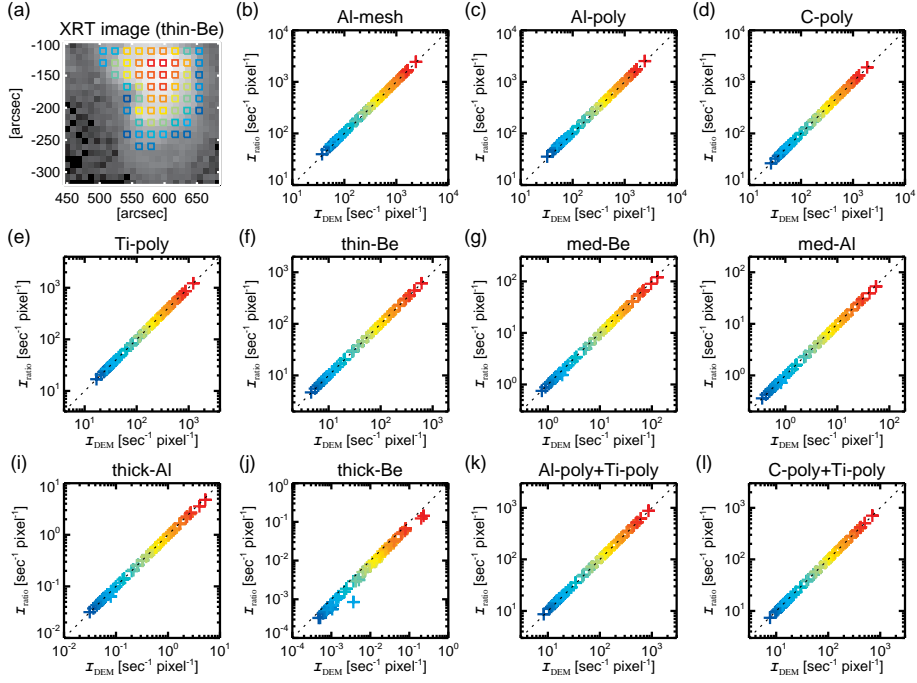}}
\vspace{2.0mm}
\caption{
Comparison between $\mathcal{I}_\mathrm{DEM}$ and $\mathcal{I}_\mathrm{ratio}$ for different regions in AR 10963.
$\mathcal{I}_\mathrm{DEM}$ and $\mathcal{I}_\mathrm{ratio}$ are calculated for each region
surrounded by a colored box in {\bf (a)}.
{\bf (b)--(l)} indicate the relation between
$\mathcal{I}_\mathrm{DEM}$ and $\mathcal{I}_\mathrm{ratio}$,
their color corresponds to the color of the box in {\bf (a)}.
The black dashed-line is the diagonal along which $\mathcal{I}_\mathrm{DEM}$ equals $\mathcal{I}_\mathrm{ratio}$.
}
\label{fig:confirmationA}
\end{figure}

Figure~\ref{fig:DEM} shows the DEMs of AR 10963 derived from EIS and XRT data
with the same method described in \inlinecite{win12}.
In this analysis, we use 7 lines observed with EIS to determine the lower temperature component
of the DEM and the data observed with 2 XRT filters to constrain the higher temperature component (see Table~\ref{tbl:DEM}).
Note that we only use the XRT filters well calibrated in Paper~I, namely, Ti-poly and thin-Be filters,
in this DEM analysis.
Each colored line in Figure~\ref{fig:DEM} shows the DEM of a region indicated by
each colored box in Figure~\ref{fig:confirmationA} (a). The size of each color box
is 9 $\times$ 9 pixels in the XRT image, which corresponds to 9.27$''$ $\times$ 9.27$''$.

Let us now check the validity of our method using the DEM.
First, we calculate the X-ray intensity observed with each focal-plane analysis filter, $\mathcal{I}_\mathrm{DEM}$,
for each observed XRT region which has the DEM in Figure~\ref{fig:DEM}.
Using the temperature response, $F(T)$, of XRT (see Figure~\ref{fig:response}),
we can calculate $\mathcal{I}_\mathrm{DEM}$ for each filter as follows:
\begin{equation}
\mathcal{I}_\mathrm{DEM} =  \int F(T) \times DEM\left(T\right) dT  .
\label{eq:I multi}
\end{equation}

Next, we derive the filter-ratio temperature $\mathcal{T}_\mathrm{ratio}$ and
volume emission measure $\mathcal{VEM}_\mathrm{ratio}$
from $\mathcal{I}_\mathrm{DEM}^\mathrm{Ti}$ and $\mathcal{I}_\mathrm{DEM}^\mathrm{Be}$,
where $\mathcal{I}_\mathrm{DEM}^\mathrm{Ti}$ and $\mathcal{I}_\mathrm{DEM}^\mathrm{Be}$ are the calculated intensities
where XRT observes the DEMs in Figure~\ref{fig:DEM} with Ti-poly and thin-Be filters, respectively.
Using $\mathcal{T}_\mathrm{ratio}$ and $\mathcal{VEM}_\mathrm{ratio}$,
we calculate the intensity $\mathcal{I}_\mathrm{ratio}$
in the same way as described in step~(II) of Section~\ref{sec:calibration}.

In Figure~\ref{fig:confirmationA} (b)--(l), we compare $\mathcal{I}_\mathrm{DEM}$ and $\mathcal{I}_\mathrm{ratio}$
for each filter.
In these plots, we see that $\mathcal{I}_\mathrm{ratio}$ is almost identical to $\mathcal{I}_\mathrm{DEM}$
for the various regions for which the DEMs are shown in Figure~\ref{fig:DEM}.
In Section~\ref{sec:calibration}, we calibrated the filter thicknesses with the intensity $I_\mathrm{exp}$
estimated from $T_\mathrm{ratio}^\mathrm{Be/Ti}$ and ${VEM}_\mathrm{ratio}^\mathrm{Be/Ti}$,
though the active region should contain multi-temperature structures.
Based on the above consistency between $\mathcal{I}_\mathrm{DEM}$ and $\mathcal{I}_\mathrm{ratio}$,
the quantity $I_\mathrm{exp}$ (corresponding to $\mathcal{I}_\mathrm{ratio}$) should
represent sufficiently well the expected X-ray intensity from the region
(corresponding to $\mathcal{I}_\mathrm{DEM}$) for calibrating the filter thicknesses.
Hence, we conclude that the 
calibration method using the filter ratio, discussed in Section~\ref{sec:calibration},
is appropriate for a multithermal structure observed by XRT.

We note that since the cross-calibration between XRT and EIS has not been completed,
we use the DEMs derived from EIS and XRT data (Figure~\ref{fig:DEM}) as the current best estimate values.
Therefore, in this work, we do not calibrate the XRT filters using such DEMs
as an absolute reference.

% ************************************************************************ SUB-SECTION
\subsection{Confirmation Taking a Different Quiescent Active Region}
\label{sec:confirmationB}

% ........................................................................ FIGURE
\begin{figure}
\centerline{\includegraphics[width=12.0cm,clip=]{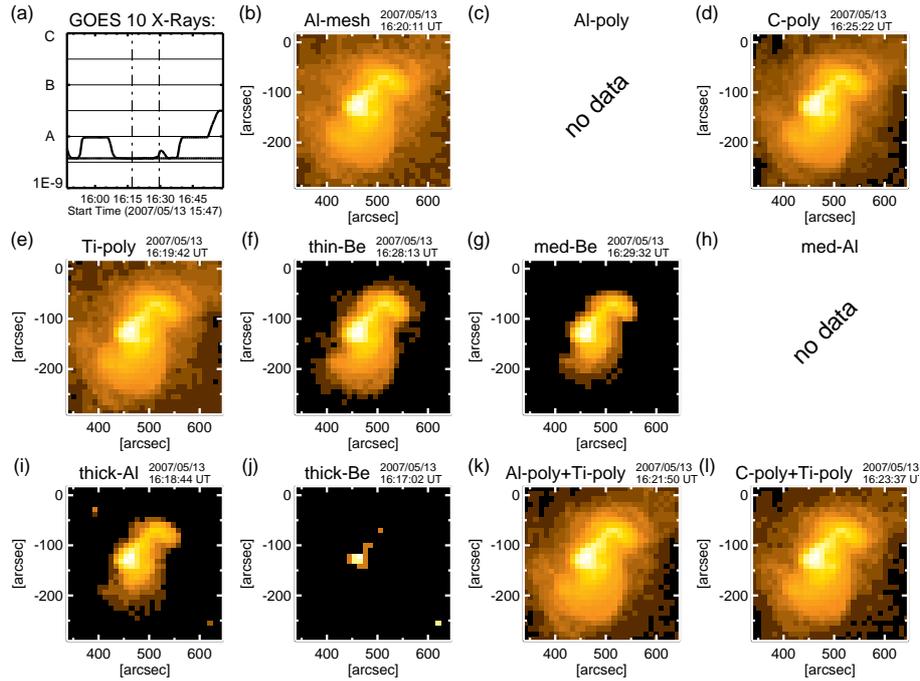}}
\vspace{2.0mm}
\caption{
Datasets of the quiescent active region (AR 10955) used to confirm the calibration results.
These images were taken by XRT with the multiple filters from 16:17 to 16:29 UT on 2007 May 13.
The meaning of each panel is the same as in Figure~\ref{fig:datasets}.
}
\label{fig:confirmationB datasets}
\end{figure}

% ........................................................................ FIGURE
\begin{figure}
\centerline{\includegraphics[width=12.0cm,clip=]{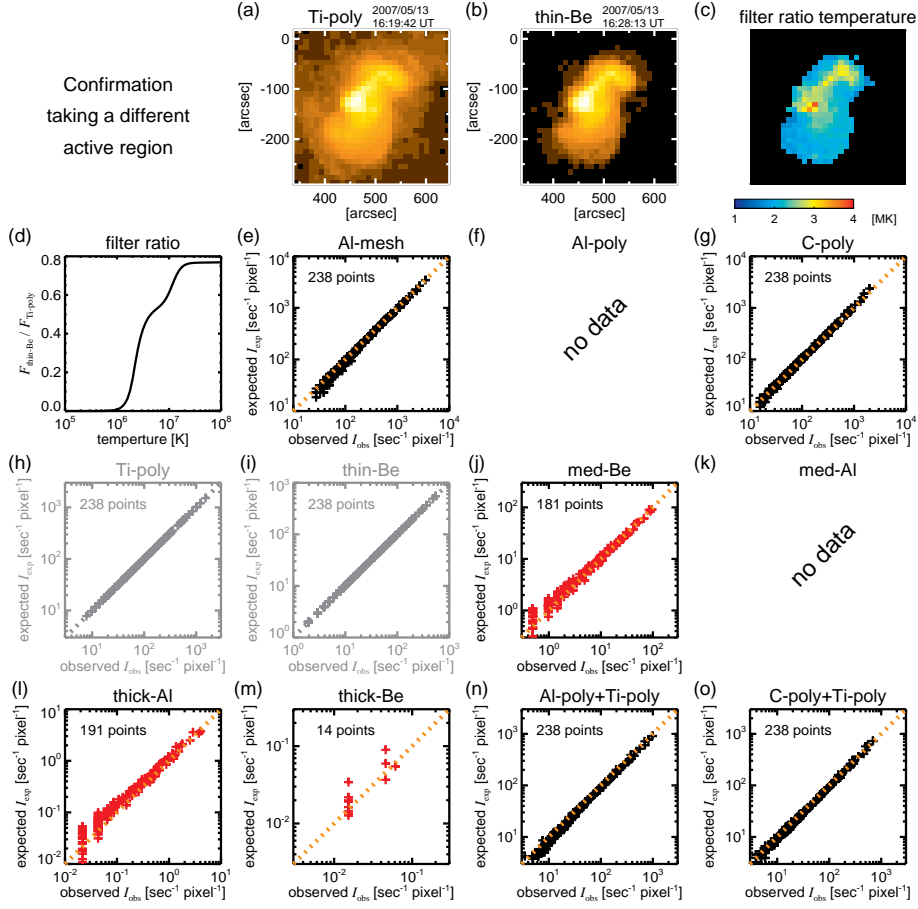}}
\vspace{2.0mm}
\caption{
Calibration results using $T_\mathrm{ratio}^\mathrm{Be/Ti}$ and ${VEM}_\mathrm{ratio}^\mathrm{Be/Ti}$ derived from
Ti-poly and thin-Be filter-pair data of AR 10955.
The meaning of each panel and symbol is the same as in Figure~\ref{fig:cal result}.
}
\label{fig:confirmationB}
\end{figure}

Next, we check the results of the calibration
taking a different quiescent active region, AR 10955 (see Figure~\ref{fig:confirmationB datasets}).
Along this period, GOES X-ray intensity level was stable as shown in Figure~\ref{fig:confirmationB datasets} (a).
The ``standard candle" filters used here are the same as those used in Section~\ref{sec:calibration}
(Ti-poly and thin-Be filters).
Images with Al-poly and med-Al filter were not taken in this period.
Figure~\ref{fig:confirmationB} shows that
$I_\mathrm{exp}$ is consistent with $I_\mathrm{obs}$ for all filters which were used to observe
this active region (black plus signs in the panels are based on the calibration results in Paper~I,
while red plus signs are based on the calibration made in this paper).
This result suggests that the thicknesses determined in Section~\ref{sec:calibration}
are also valid for this active region.

% ************************************************************************ SUB-SECTION
\subsection{Summary of the Calibration Results}
\label{sec:confirmations}

In Section~\ref{sec:confirmationA}, we demonstrate the validity of using $I_\mathrm{exp}$
estimated from $T_\mathrm{ratio}^\mathrm{Be/Ti}$ and ${VEM}_\mathrm{ratio}^\mathrm{Be/Ti}$
for the calibration of the filter thicknesses.
In Sections~\ref{sec:confirmationB}, we confirm the consistency of the results using data from
a different active region.

% ........................................................................ FIGURE
\begin{figure}
\centerline{\includegraphics[width=5.0cm,clip=]{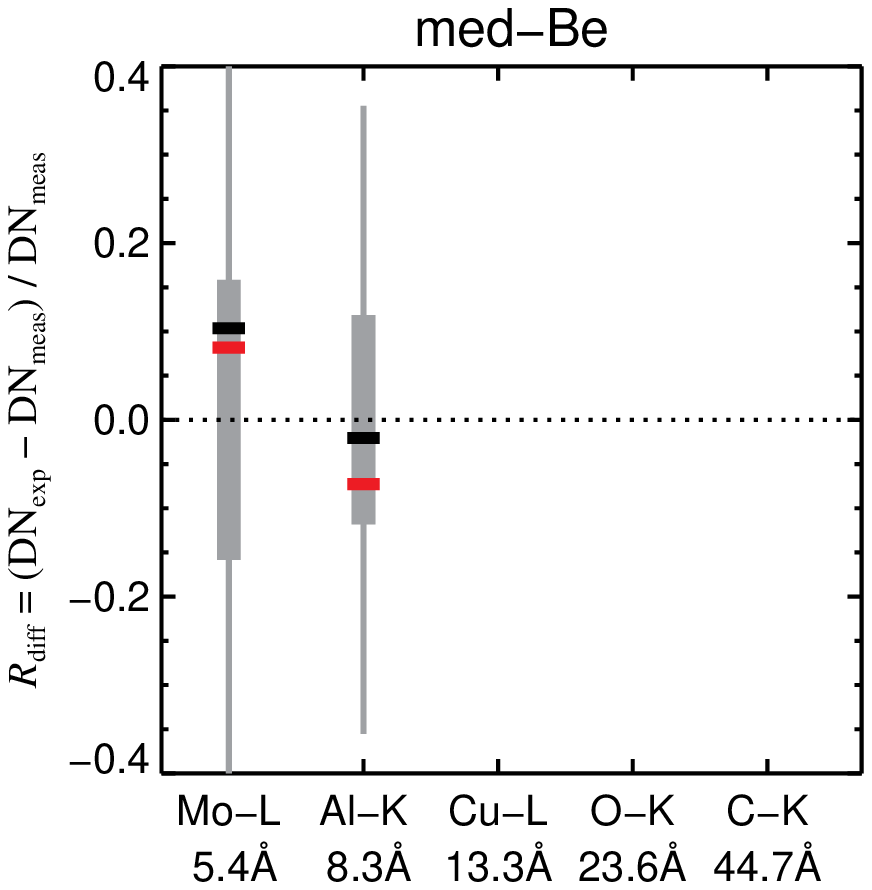}
            \hspace*{0.0cm}
            \includegraphics[width=5.0cm,clip=]{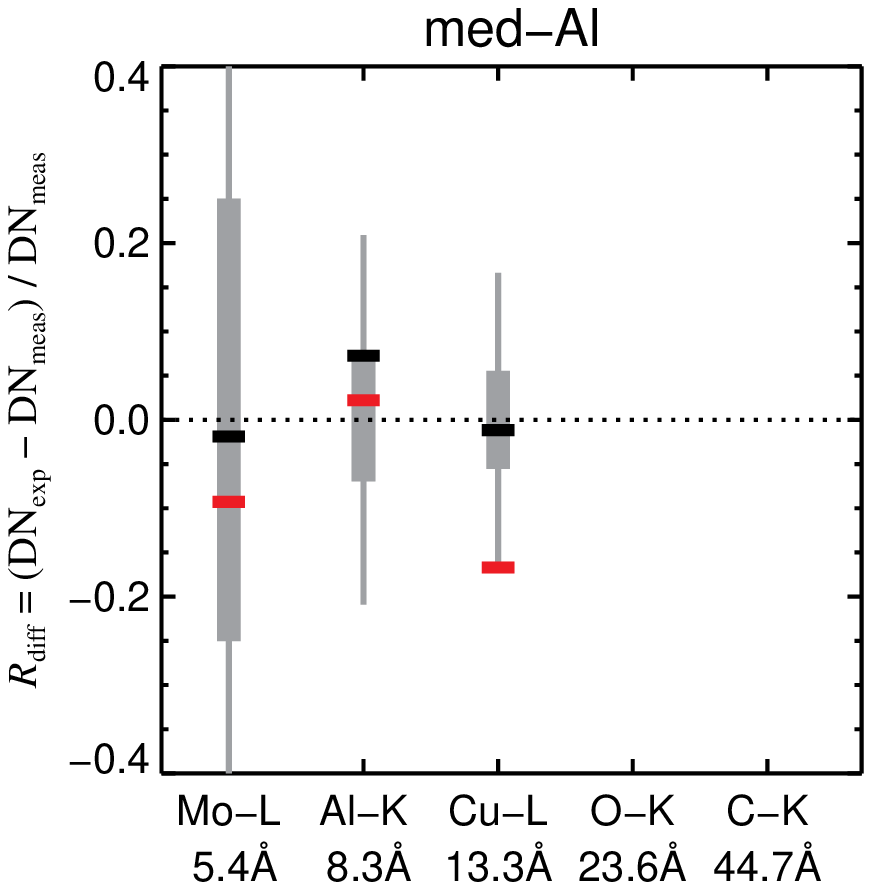}
           }
\centerline{\includegraphics[width=5.0cm,clip=]{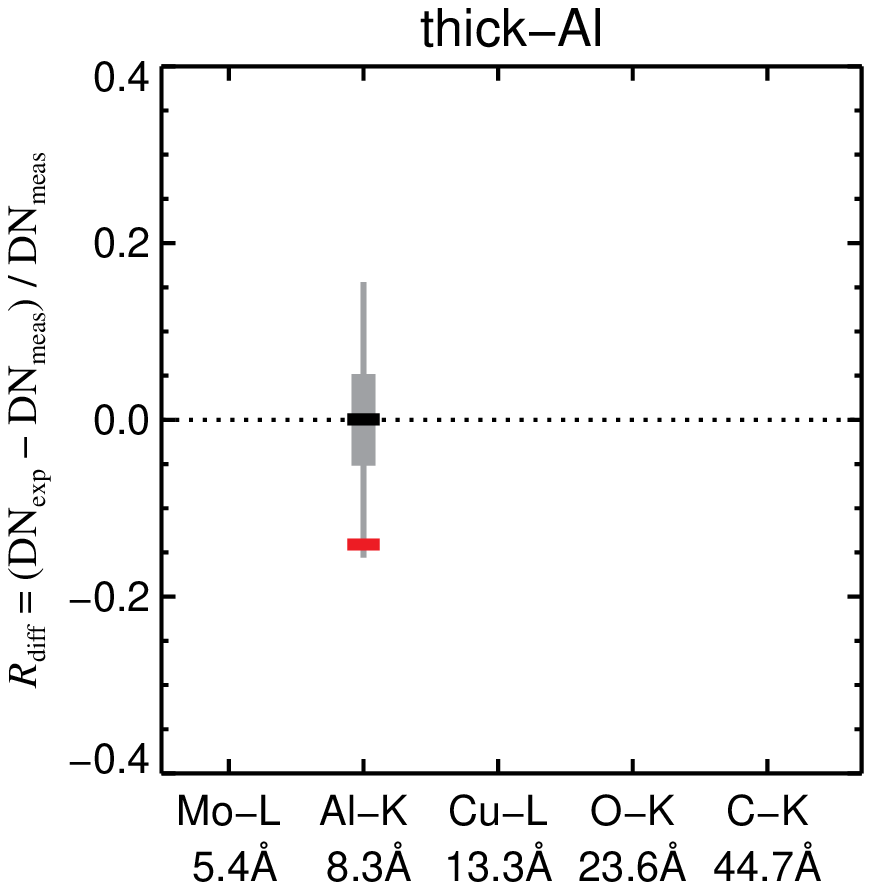}
            \hspace*{0.0cm}
            \includegraphics[width=5.0cm,clip=]{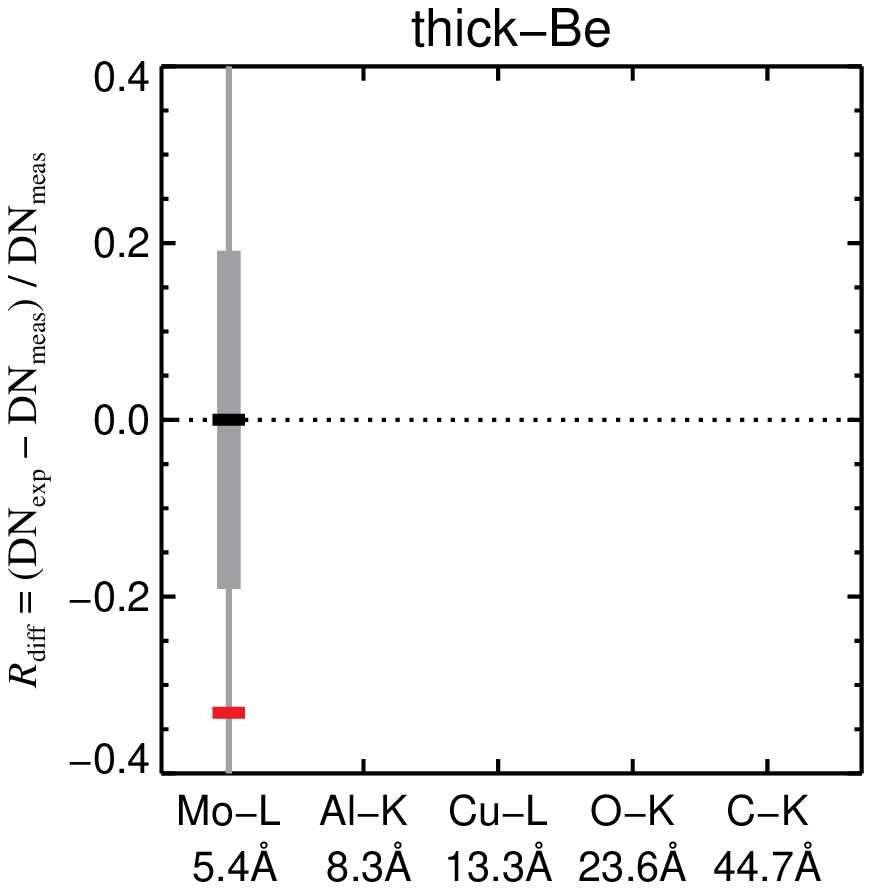}
           }
\vspace{2.0mm}
\caption{
Comparison between the measured data numbers, $\mathrm{DN}_\mathrm{meas}$, by the XRT CCD
in the ground-based end-to-end test,
and the expected data numbers, $\mathrm{DN}_\mathrm{exp}$, based on the calibration results in this paper.
These plots are similar to Figures~21 and 22 of Paper~I.
The minus ($-$) signs indicate the differences defined as
$R_\mathrm{diff} \equiv (DN_\mathrm{exp}-DN_\mathrm{meas})/DN_\mathrm{meas}$.
The red and black minus ($-$) symbols indicate $R_\mathrm{diff}$ based on
the calibration results in this paper and Paper~I, respectively.
The names of the examined filters are indicated in each top label.
The labels in the abscissa show the X-ray emission lines and their wavelength, used in the end-to-end test.
The thick and thin gray bars are the $1\sigma$ and $3\sigma$ error bars, respectively.
}
\label{fig:e2e}
\end{figure}

Now, let us check whether the filter thicknesses, determined in this work, are consistent with
the transmissivity measurements of the filters performed in the ground-based end-to-end test at XRCF
(see Appendix~A of Paper~I for the details of the end-to-end test).
Figure~\ref{fig:e2e} shows the difference defined as
$R_\mathrm{diff} \equiv (\mathrm{DN}_\mathrm{exp}-\mathrm{DN}_\mathrm{meas})/\mathrm{DN}_\mathrm{meas}$,
where $\mathrm{DN}_\mathrm{exp}$ is the expected data numbers obtained from the incident X-ray intensity spectrum
monitored by a flow proportional counter with the XRT filter thickness
and $\mathrm{DN}_\mathrm{meas}$ is the measured DN by the XRT CCD.
If the incident X-ray intensity is large enough to have a good signal-to-noise ratio,
the spectrum of incident X-rays can be well monitored and
the XRT filter thickness can be determined with high precision.
This means ideally that the value of $R_\mathrm{diff}$ is zero.
In reality, however, $R_\mathrm{diff}$ has error bars derived from
the photon noise of the X-rays detected by the XRT CCD and
the uncertainties of the incident X-ray spectrum monitored by the flow proportional counter,
as shown by the gray bars in Figure~\ref{fig:e2e}.
The values of $R_\mathrm{diff}$ calculated with the filter thickness determined in this work
are indicated by red minus signs in Figure~\ref{fig:e2e}, and they are located within $3 \sigma$ error bars
shown with thin gray bars.
This suggests that our calibration results do not contradict the ground-based end-to-end tests.

Hence, here we conclude that the thicknesses of the X-ray focal-plane analysis filters
including the thicker filters are well calibrated.

% ************************************************************************ SECTION
\section{Coronal-Temperature-Diagnostic Capability of XRT Based on the Calibration Results}
\label{sec:summary}

% ........................................................................ FIGURE
\begin{figure}
\centerline{\includegraphics[width=12.0cm,clip=]{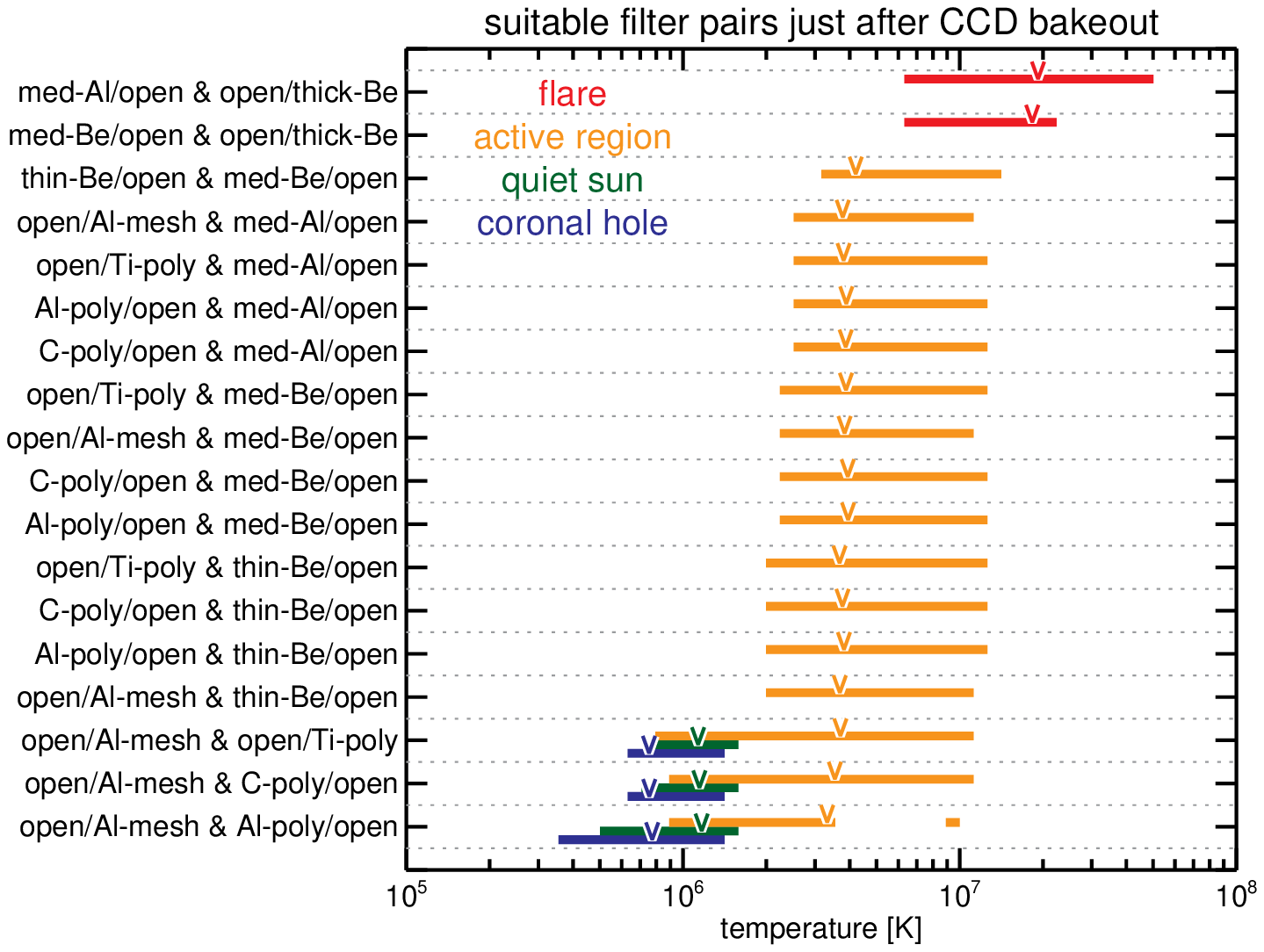}}
\centerline{\includegraphics[width=12.0cm,clip=]{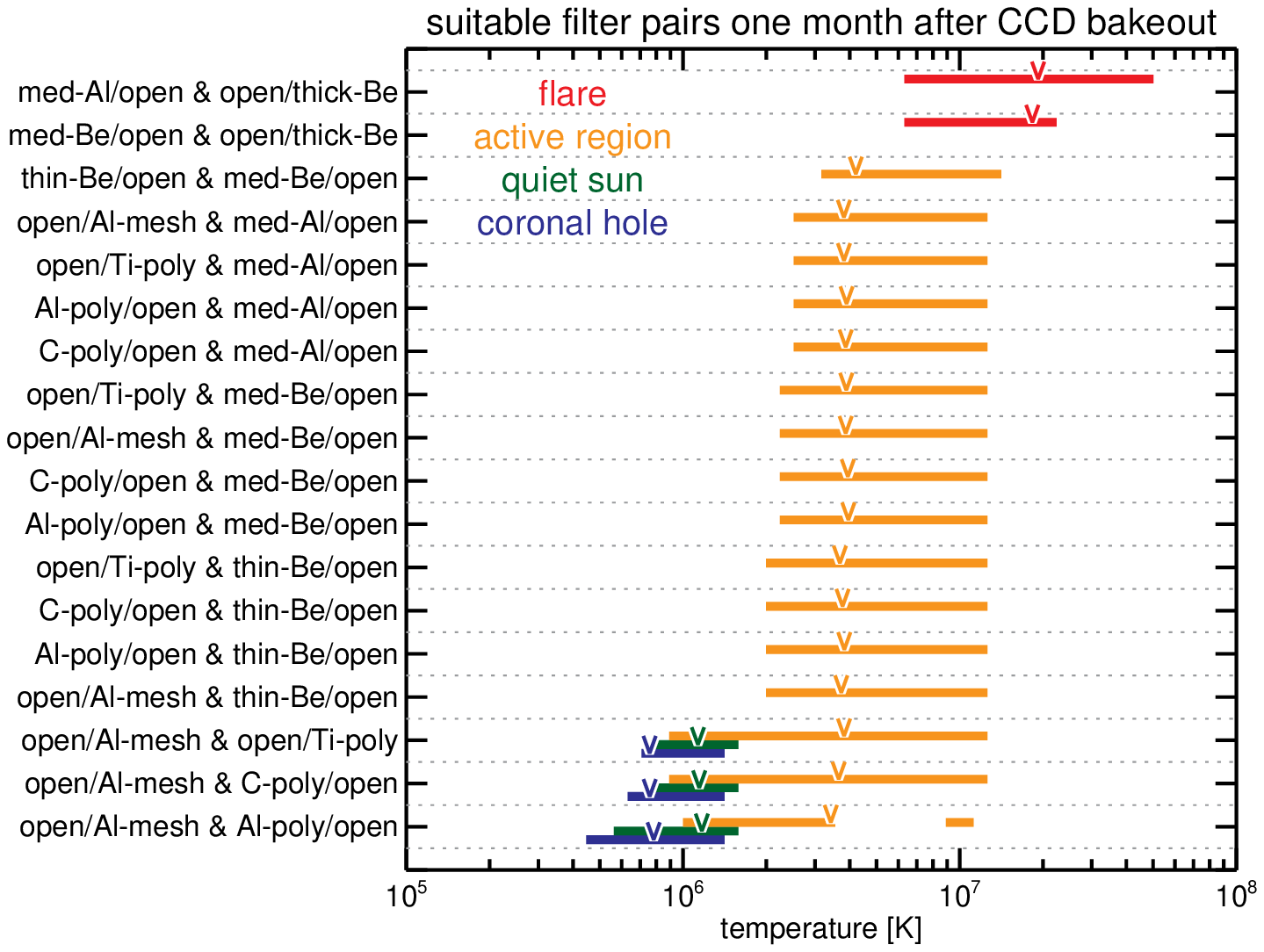}}
\vspace{2.0mm}
\caption{
Estimable temperature ranges derived using the filter-ratio method.
Top and bottom panels show examples
just after the CCD bakeout and one month after the CCD bakeout, respectively.
We assume that 800~{\AA} of laminar contaminant accumulated on the CCD during one month after the CCD bakeout.
The red, orange, green, and blue bars show the estimable temperature ranges
for flares, active regions, quiet Sun, and coronal holes, respectively.
}
\label{fig:suitable filter pair}
\end{figure}

Figure~\ref{fig:suitable filter pair} is
the updated estimable temperature ranges derived using the filter-ratio method for various coronal structures, namely,
flares, active regions, quiet Sun and coronal holes, based on the calibration results found in this work.
The details of how to obtain the estimable temperature ranges are described in Section~5.2 in Paper~I.
On the basis of Figure~\ref{fig:suitable filter pair},
we find that XRT has a capability to diagnose the whole coronal-temperature range from
less than 1~MK to more than 10~MK, {\it i.e.}, from coronal holes to flares.

% ........................................................................ FIGURE
\begin{figure}
\centerline{\includegraphics[width=12.0cm,clip=]{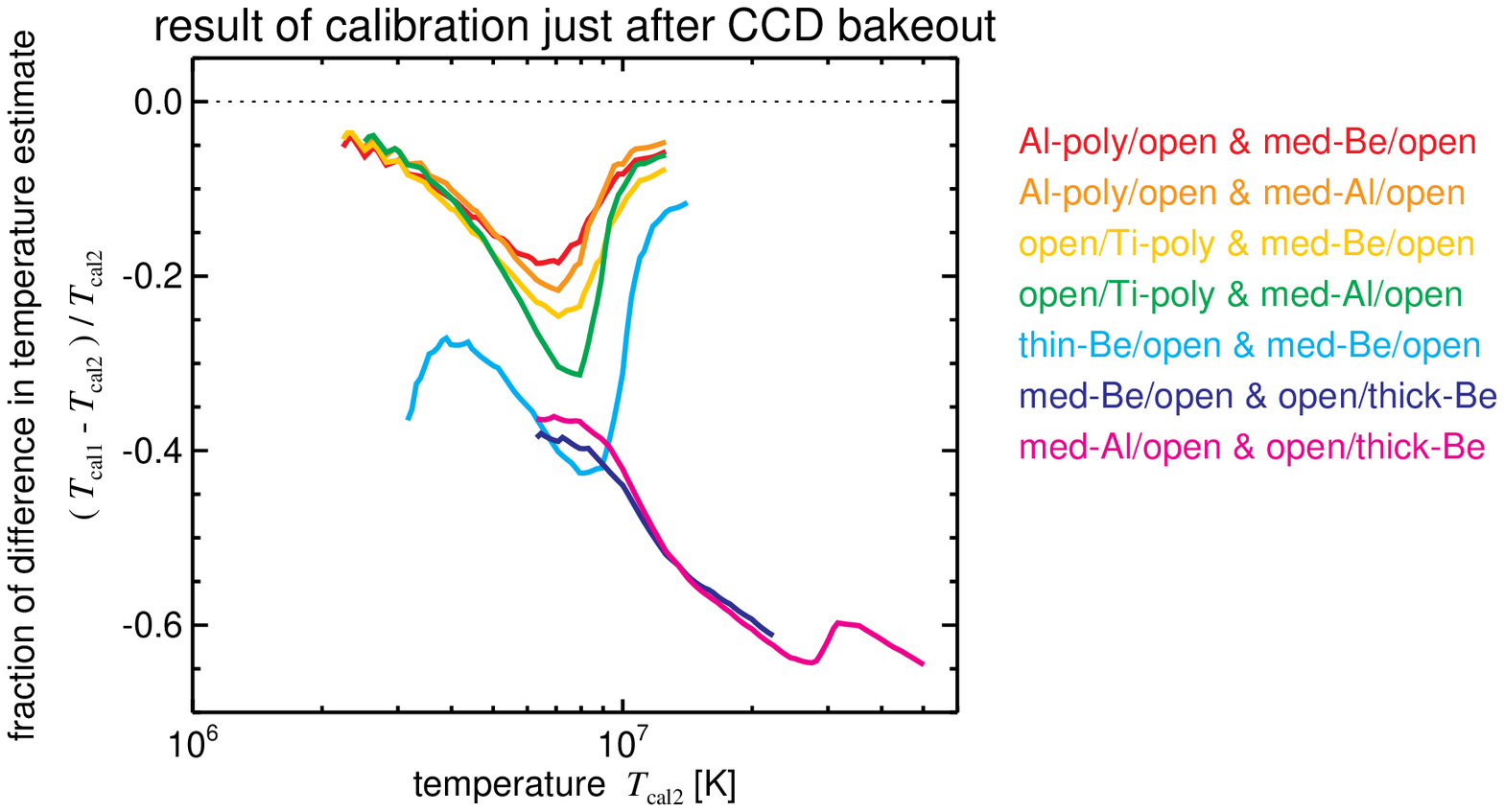}}
\caption{
Difference in the derived filter-ratio temperature based on the calibration results
found in Paper~I (cal1) and in this paper (cal2).
}
\label{fig:diff}
\end{figure}

% ........................................................................ FIGURE
\begin{figure}
\centerline{\includegraphics[width=12.0cm,clip=]{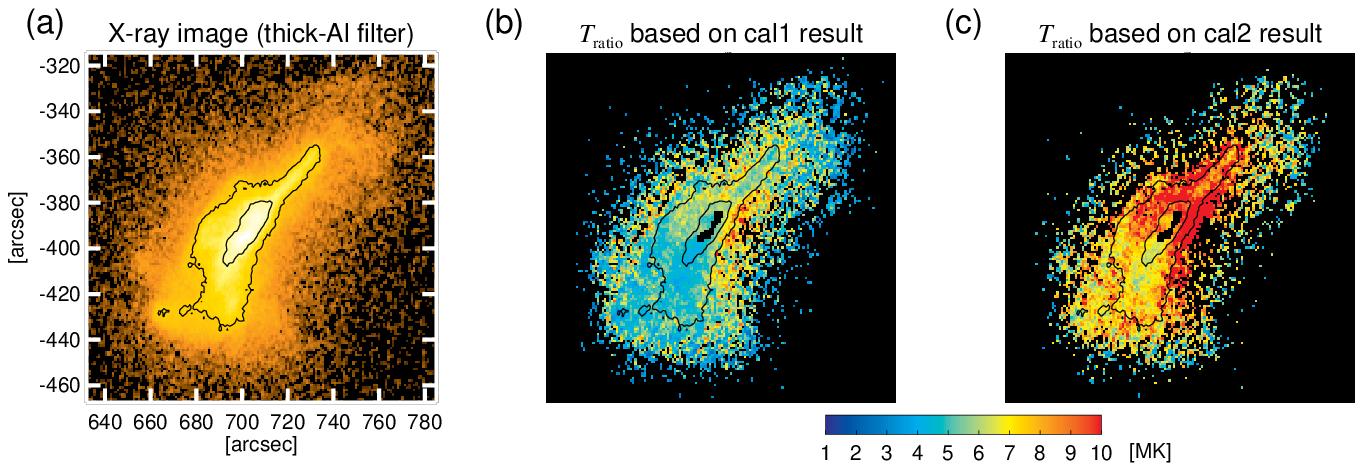}}
\caption{
Derived filter-ratio temperature for the M-class flare observed at 6:30 on 2011 June 7 in AR~11226.
{\bf (a)} is the X-ray image taken with the thick-Al filter.
{\bf (b)} and {\bf (c)} show the filter-ratio temperatures derived with the filter pair of thick-Al and
thick-Be based on the calibration results found in Paper~I (cal1) and in this paper (cal2), respectively.
The contours indicate the area where the count rate in the thick-Al image (shown in {\bf (a)}) is
800 and 3200 [DN sec$^{-1}$ pixel$^{-1}$], respectively.
}
\label{fig:flare}
\end{figure}

Finally, we summarize the difference between the calibration results in Paper~I
and in this work. The updated thicknesses of med-Be, med-Al, thick-Al and thick-Be filters
(see Table~\ref{tbl:FPAF}) are larger than those in Table~1 of Paper~I.
Figure~\ref{fig:diff} shows the difference in the derived filter-ratio temperatures determined with
the filter thicknesses found in Paper~I and the ones in this paper.
When med-Be, med-Al, thick-Al and/or thick-Be filters are used for the coronal-temperature diagnostics,
the filter-ratio temperatures determined using the results in Paper~I
are lower than those derived with the results found in this paper.
Figure~\ref{fig:flare} shows an example of filter-ratio temperature analysis for an M-class flare.
Figure~\ref{fig:flare} (a) is the X-ray image observed with the thick-Al filter at the peak of GOES X-ray flux. 
Figures~\ref{fig:flare} (b)--(c) show the filter-ratio temperatures derived with the filter pair of thick-Al and
thick-Be using the results in Paper~I and those in this paper, respectively.
In Figure~\ref{fig:flare} (b), the bulk of the flare plasma is at a temperature of at most 5~MK,
but calculations using the two bands of GOES total X-ray flux ({\it i.e.}, 1--8~{\AA} and 0.5--4~{\AA})
indicate that the bulk plasma temperature is 12.8~MK.
With the filter thicknesses determined in this paper, the flare temperature is found to be
greater than 10~MK, which is consistent with GOES measurements.

We conclude that
we now have an updated set of filter thicknesses which gives consistent and reasonable filter-ratio temperatures
for coronal plasmas and which are also consistent with pre-launch measurements.
With the new set of filter thicknesses,
we can make full use of the coronal-temperature-diagnostics capability of XRT;
in particular, the diagnostic capability for high temperature plasmas has been drastically improved
with the updates of the thick filter group.

% ######################################################################## ACKNOWLEDGEMENTS
%\acknowledgements
\begin{acks}
The authors thank members of the XRT team for useful discussions and comments.
\textit{Hinode} is a Japanese mission developed and launched by ISAS/JAXA,
collaborating with NAOJ as a domestic partner, NASA and STFC (UK)
as international partners.
Scientific operation of the \textit{Hinode} mission is conducted
by the \textit{Hinode} science team organized at ISAS/JAXA.
This team mainly consists of scientists from institutes
in the partner countries.
Support for the post-launch operation is provided by
JAXA and NAOJ (Japan), STFC (UK), NASA, ESA, and NSC (Norway).
MW and KKR acknowledge support from contract NNM07AB07C through NASA to SAO.
\end{acks}

% ######################################################################## BIBLIOGRAPHY
% format of references provided by the journal (.bst)
% \bibliographystyle{spr-mp-sola}
\bibliographystyle{spr-mp-sola-cnd} %% Alternative style: no title, no concluding page. 
\bibliography{Narukage_etal_2012}

\end{article} 

\end{document}